\documentclass[a4paper,10pt]{article}

\pdfoutput=1 

\usepackage{jheppub} 

\usepackage[T1]{fontenc} 

\usepackage{amsfonts}
\usepackage{bbm}

\numberwithin{equation}{section}

\newcommand{\be}{\begin{equation}}
\newcommand{\ee}{\end{equation}}

\numberwithin{equation}{section}

\def\cA{{\cal A}}

\def\cC{{\cal C}}

\def\cW{{\cal W}}

\def\be{\begin{equation}}
\def\ee{\end{equation}}
\def\bea{\begin{eqnarray}}
\def\eea{\end{eqnarray}}
\def\ba{\begin{array}}
\def\ea{\end{array}}

\Huge
\title{\bf\large\sc On  the Casimir ${\cW\cA}_{\it{N}}$\,  algebras as the truncated $\cW_{\infty}$\,algebra }
\author{H.T. ~\"OZER}
\affiliation{Istanbul Technical University,Faculty of Science and Letters, Physics Department \\34469 Maslak Istanbul,TURKEY}

\emailAdd{ozert@itu.edu.tr}

\abstract{The complete structure of   the Casimir ${\cW\cA}_{\it{N}}$\,  algebras are shown to exist in such a way that the
Casimir ${\cW\cA}_{\it{N}}$\,algebra is a kind of truncated type of $\cW_{\infty}$\,algebra both in the primary and in the quadratic
basis, first using the associativity conditions in the basis of primary fields and second using the Miura basis coming from
the free field realization as a different basis.\,Finally one can say that the Casimir ${\cW\cA}_{\it{N}}$\,algebra is a kind
of truncated type of $\cW_{\infty}$\,algebra,so it is clear from any construction of $\cW_{\infty}$\,algebra that by
putting infinite number of fields $W_{s}$ with $s>N$ to zero we arrive at the Casimir ${\cW\cA}_{\it{N}}$\,algebra.}

\begin{document}
\maketitle


\section{Introduction}
\label{sec:intro}
Although a wide variety of $\cW_{\it{N}}$\,algebras was discovered since the last quarter century,
nowadays we still do not have an explicit form for the Operator Product Expansions(OPEs) between primary fields and
the structure constants for sufficiently large values of $\it{N}$, therefore the higher spin extensions of
the conformal symmetries is still an open problem.

Conformal symmetry play an important role in the study of two - dimensional conformal field theories and thus
it founds striking applications in string theory \cite{Witten} and in the study of critical phenomena
in statistical physics \cite{Itzykson}, as well as in mathematics \cite{bpz84}. Underlying symmetry algebra of
this symmetry is Virasoro algebra which appears naturally in two - dimensional conformal field theories.\,
The idea to extend Virasoro algebra with the introduction of higher conformal spin generators
is also seem to be relevant in these theories.\ A seminal type of these extensions is the so - called
$\cW_{N}$\,algebras and Virasoro algebra is a $\cW_{\it{2}}$\,algebra within this framework. First example of
the extended symmetry algebra,$\cW_{\it{3}}$, was constructed by Zamolodchikov in \cite{Zamolodchikov:1985wn,fatzam}.
This algebra was obtained by extending the Virasoro algebra by addition of spin-3 conformal field.
This algebras have been studied dealing with the classification and also construction from a variety
of view points \cite{HamadaTakao,Zhang,KauschWatts,Blumenhagenatall,Hansoy},\,as well as from the Casimir
${\cW\cA}_{\it{N}}$\,algebras \cite{Thierry,Bais,Ozer} point of view.

Another intriguing $\cW$\,algebra very closely related to $\cW_{\it{N}}$\,algebra  is  $\cW_{\infty}$\,family,
that is,$\cW$\,algebra can be interpreted in the context of a $\cW_{\infty}$\,type algebra
\cite{Pope:1989ew,Pope:1989sr,Pope:1990kc,Lu:1991pe}.\,Such an algebra is based on an infinite number
of higher - spin operators with spin $s\geq 1$, or 2 and multiplicity 1. One emphasize here that the authors were
studying $\cW_{\infty}$ for special value of parameters where the algebra linearizes. The first signs of connection between
$\cW_{\infty}$ and $\cW_{\it{N}}$ is in the seminal Ref. \cite{Lu:1991pe}. From today's point of view
one can say that $\cW_{\infty}$ is two-parametric family algebra \cite{Gaberdiel:2011wb,Gaberdiel:2011zw,Gaberdiel:2012ku}.
The connection between one - parametric algebra $\cW_{\it{N}}$ and two-parametric family $\cW_{\infty}$ is similar to the
construction of the higher spin algebra $\mathfrak{hs}(\lambda)$ in Vasiliev theory \cite{Vasiliev:1999ba}.From today's
point of view, the main idea of related studyings  is  first to understand $\cW_{\infty}$ and truncate it to
$\cW_{\it{N}}$ for any $\it{N}$.There are some explicit results for quantum case as considered in\cite{Prochazka:2014gqa}
where the author focuses  on the construction of $\cW_{\infty}$ in the primary basis and also
in the Miura basis\,\cite{Drinfeld:1984qv,Fateev:1987zh,Bilal:1989db,Fateev:1990,Bilal:1991trieste,Bouwknegt:1992wg}
which is different from the primary bases.

We also emphasize here that a wide variety of $\cW_{\it{N}}$\,algebras was discovered as expressed above, but nowadays we still do not have
an explicit form for the Operator Product Expansions between primary fields and  the structure constants for sufficiently
large values of $\it{N}$. Therefore the main purpose of our work is  to try to fill in this gap by finding the complete structure of
the Casimir ${\cW\cA}_{\it{N}}$\, algebras for sufficiently large values of $\it{N}$ and to establish a method in this
direction in the Casimir ${\cW\cA}_{\it{N}}$\,algebra point of view as was done in\cite{Ozer}.

The paper is organized as follows:\, In section 2 we consider that $\cW_{\it{N}}$\,algebra is a kind of truncated type of $\cW_{\infty}$\,algebra,
then,\, we construct the most general $\cW_{\it{N}}$\,algebra in the primary basis,and explain how the different structure constants
can be determined recursively from the Jacoby identity.\,For this,we first count  primary fields
in the field content of the $\cW_{\it{N}}$ just as in the field content of the $\cW_{\infty}$,that is,\,using the character formula
for $\cW_{\infty}$, one can get a counting function which counts the Virasoro primaries for $\cW_{\infty}$,\, so one can get truncated form of
this character formula which counts the Virasoro primaries for $\cW_{\it{N}}$. The method is rather straightforward : one begin with
the stress-\, energy tensor and extends the algebra by addition of the higher-\,spin primary fields $W_{s}(z)$ with spin $s=3,4,...,N$.\,
These primary fields generate all other fields by using the conditions of the associativity  of the OPEs.

In the next part,\,we made use of construction known as Miura transformation with Feigin - Fuchs type of free massless scalar fields, as
a different starting point.\,It is seen that this gives us the possibility to exploit a relation between $\cW_{\it{N}}$- algebras
and $\cA_{{N-1}}$-Lie algebras.\,This brings us to the fact that one can define the primary fields of $\cW\cA_{\it{5}}$ Casimir algebra
as was done in\cite{Ozer}.\,So we proceed similarly to what we did in the first part and use the associativity conditions to compute the OPEs.
In this way we can make comparison between results of the first and second parts of the article.\, In the Appendix we give the explicit OPEs
that overlap with either $\cW_{\it{N}}$- algebras having structure constants(2.13) in section 2 and the Casimir ${\cW\cA}_{\it{N}}$\,
algebras having structure constants(3.17)in the section 3.\,All these have been possible with a dense application of
Mathematica Package program \cite{WolframMath}.


\section{ $\cW_{\it{N}}\,$algebra in the primary basis}
In this section we will study the $\cW_{\it{N}}$  algebra in the basis of the Virasoro primary fields. To get the $\cW_{\it{N}}$ algebra,
we will extend the Virasoro algebra generated by stress-energy tensor $T(z)$
\be
\begin{split}
& T(z_1)T(z_2)\,\sim \,{{c\over 2}\over{z_{12}^{4}}}\,+\, {2\,T\over{z_{12}^{2}}}\, + \,{\partial T\over{z_{12}}}
\end{split}
\ee
by higher spin primary generators $W_{s}(z)$ of spin $s=3,4,...,N$:
\be
\begin{split}
& T(z_1)W_{s}(z_2)\,\sim \,{s\,W_{s}\over{z_{12}^{2}}}\, + \,{\partial W_{s}\over{z_{12}}}\, .
\end{split}
\ee
We will usually denote  the Virasoro generator by $T(z)\equiv W_{\it{2}}(z)$ and $z_{12}=z_{1}-z_{2}$ .
The algebraic structure will be fixed by imposing the conditions of associativity of the operator product algebra.
We do not start  by reviewing the construction and algebraic properties of the normal ordering in the radial
quantization since  this formalism is standard enough. A large part of this article  was carried out with this
formalism by using the Mathematica package $\bf{OPEdefs}$ by Kris Thielemans \cite{Thielemans:1991uw}
to find all structure constants of related algebra.We also emphasize here that the power of the Thielemans'
 package cannot be denied.


\subsection{Conformal Bootstrap and $\cW_{\it{N}}\,$algebra in the primary basis }

We can first define the field content of  $\cW_{N}$ and  apply the basic proporties of operator product expansions
(OPE) to compute the OPE of $\cW_{\it{N}}$.


\subsubsection{ Field content of  $\cW_{\it{N}}$ from $\cW_{\infty}$\,algebra and counting $W_{s}(z)$  primaries}

One can say that $\cW_{\it{N}}$\,algebra is a kind of truncated type of $\cW_{\infty}$\,algebra,so it is clear from any construction
of $\cW_{\infty}$\,algebra that by putting infinite number of fields $W_{s}$ with $s>N$ to zero we arrive at $\cW_{\it{N}}$\,algebra.
The definition of field content of  $\cW_{\it{N}}$ will be  our starting point in this paper. As field content,
this algebra consist of two part; the first part have simple primary fields $W_{s}(z)$ of spin $s=3,4,...,N$, and all
other fields of $\cW_{\it{N}}$ will be obtained by taking derivatives an normal ordered products of these fields as composite primary fields.

\noindent Using the character formula
\be
\begin{split}
\chi[q]=q^{-{c\over 24}} \prod_{s=2}^{\infty}\prod_{j=s}^{\infty}{1\over{1-q^{j}}}
\end{split}
\ee
\noindent for the vacuum representation of  $\cW_{\infty}$\,algebra, we can write down the function $P[q]$ and $P_{h}$ is the integer counting
the number of Virasoro primaries of conformal dimension h
\be
\begin{split}
P[q]=\sum_{h=0}^{\infty}P_{h}q^h
\end{split}
\ee
that counts the Virasoro primaries  for $\cW_{\infty}$\,algebra.Then we find a formula for $P[q]$
\begin{equation}
\label{primarycounting}
P(q) = q + (1-q) \prod_{s=3}^{\infty} \prod_{j=s}^{\infty} \frac{1}{1-q^j} \simeq 1 + q^3 + q^4 + q^5 + 2q^6 + 2q^7 + 5q^8 + 6q^9 + 11q^{10} + 14q^{11} + 26q^{12} + \cdots
\end{equation}
To end this,one can use the more detailed counting function
\be
\begin{split}
q + (1-q)\prod_{s=2}^{\infty}\prod_{j=s}^{\infty}{1\over{1-w_{s}q^{j}}}
\end{split}
\ee
shows what $W_{s}$ primaries create the composite one (with $w_{2}=1$). For example, at order $10$ in $q$ we find the cofficient
\begin{equation}
w_{10} + w_3 w_7+ w_4 w_6+ w_5 w_5 + w_3 w_3 w_4 + w_4 w_5 + w_3 w_6 + w_3 w_4 + w_3 w_5 + w_4 w_4 + w_3 w_3
\end{equation}
which is simply reflected by the last two lines of the table below.

\begin{center}
\begin{tabular}{|c|c|c|c|}
\hline
 spin & count& even composite  primary field                        & odd composite  primary field                                                       \\
\hline
0 & 1 & $\mathbbm{1}$                                                           & -                                                                      \\
3 & 1 &  -                                                    & $W_3$                                                                                    \\
4 & 1 &$W_4$                                                         & -                                                                               \\
5 & 1 &-                                                      &$W_5$                                                                                      \\
6 & 2 &$W_6,[W_3 W_3]$                                                              &-                                                                    \\
7 & 2 &-                                                           & $W_7,[W_3 W_4]$                                                                                    \\
8 & 5 & $W_8, [W_3 W_5],\partial^{2}[W_3 W_3], [W_4 W_4]$                                         &$\partial[W_3 W_4]$                                   \\
9 & 6 & $\partial[W_3 W_5]$&  $W_9, [W_3 W_6], [W_4 W_5], [W_3 W_3 W_3]$                                                           \\
10& 11&$W_{10},[W_3 W_7],[W_4 W_6],[W_5 W_5]$ &  $\partial^{2}[W_3 W_4],\partial[W_4 W_5],\partial[W_3 W_6] $     \\
  &   & $,[W_3 W_3 W_4],\partial^{2}[W_3 W_5]$                                                           & $\partial^{3}[W_3 W_4]$     \\
  &   &  $,\partial^{2}[W_4 W_4],\partial^{4}[W_3 W_3] $                                                                              &                                                                           \\
11&14 &  $\partial[W_3 W_3 W_4],\partial[W_4 W_6],\partial^{3}[W_3 W_5]$    &  $W_{11}, [W_3 W_8],[W_4 W_7]$                                    \\
  &   &    $,\partial^{2}[W_3 W_3 W_3],\partial[W_3 W_7]$                       &$ ,[W_5 W_6],[W_3 W_3 W_5]$                                                            \\
  &   &                                                                          & $,[W_3 W_4 W_4],\partial^{2}[W_3 W_6]$                                   \\
  &   &                                                                          &  $ ,\partial[W_4 W_5],\partial^{4}[W_3 W_4]$                                                                                      \\
\hline
\end{tabular}
\end{center}

Here these brackets are used symbolically to denote the composite primary fields.We see that from the table schematically all the composite fields
 are obtained by taking derivatives and normal ordered products of single primary fields .

\subsubsection{ Conformal Bootstrap and an ansatz for OPEs of $\cW_{\it{5}}$\,algebra with associativity }

We  emphasize here that $\cW_{\it{5}}$\,algebra is a kind of truncated type of $\cW_{\infty}$\,algebra,so it is clear from any construction
of $\cW_{\infty}$\,algebra that by putting infinite number of fields $W_{s}$ with $s>5$ to zero we arrive at $\cW_{\it{5}}$\,algebra.We can summarize
all field content of $\cW_{\it{5}}$\,algebra by the following table:
\begin{center}
\begin{tabular}{|c|c|c|c|}
\hline
 spin & count& even composite  primary field                        & odd composite  primary field                                                       \\
\hline
0 & 1 & $\mathbbm{1}$                                                           & -                                                                      \\
3 & 1 &  -                                                    & $W_3$                                                                                    \\
4 & 1 &$W_4$                                                         & -                                                                               \\
5 & 1 &-                                                      &$W_5$                                                                                      \\
6 & 1 &$[W_3 W_3]$                                                              &-                                                                    \\
7 & 1 &-                                                           & $[W_3 W_4]$                                                                                    \\
8 & 4 & $[W_3 W_5],\partial^{2}[W_3 W_3], [W_4 W_4]$                                         &$\partial[W_3 W_4]$                                   \\
9 & 4 & $\partial[W_3 W_5]$&  $[W_4 W_5], [W_3 W_3 W_3],\partial^{2}[W_3 W_4]$                                                             \\
\hline
\end{tabular}
\end{center}
To get this table we use the following truncated  form of  equation(2.6)
\be
\begin{split}
q + (1-q)\prod_{s=3}^{5}\prod_{j=s}^{\infty}{1\over{1-w_{s}q^{j}}}
\end{split}
\ee
which shows what $W_{s}$ primaries create the composite one for $\cW_{\it{5}}$\,algebra . For example, at order $8$ in $q$ we find the cofficient
\begin{equation}
w_3 w_5+ w_3 w_3 + w_4 w_4  + w_3 w_4
\end{equation}
Let me start by using $\bf{OPEdefs}$ package to find all structure constants $\cC_{ij}^{k}$ of the $\cW_{\it{5}}$\,algebra following from associativity of the OPE
algebra.Technically, we will use $\bf{OPEconf}$ packace as an extension of $\bf{OPEdefs}$. We can write an ansatz symbolically to represent OPE algebra of the primary
fields.So it can be written as:
\be
\begin{split}
& W_{3}W_{3}\,\,\sim \,\, {{c}\over {3}}\mathbbm{1} \,+\,\cC_{33}^{4}W_{4}\\
& W_{3}W_{4}\,\,\sim \,\, \cC_{34}^{3}W_{3} \,+\,\cC_{34}^{5}W_{5}\\
& W_{4}W_{4}\,\,\sim \,\, {{c}\over {4}}\mathbbm{1} \,+\,\cC_{44}^{4}W_{4}\,+\,\cC_{44}^{[33]}[W_{3}W_{3}]\\
& W_{3}W_{5}\,\,\sim \,\, \cC_{35}^{4}W_{4} \,+\,\cC_{35}^{[33]}[W_{3}W_{3}]\\
& W_{4}W_{5}\,\,\sim \,\, \cC_{45}^{3}W_{3} \,+\,\cC_{45}^{5}W_{5} \,+\,\cC_{45}^{[34]}[W_{3}W_{4}] \,+\,\cC_{45}^{[34]'}\partial[W_{3}W_{4}]\\
& W_{5}W_{5}\,\,\sim \,\, {{c}\over {5}}\mathbbm{1} \,
+\,\cC_{55}^{4}W_{4}\,
+\,\cC_{55}^{[33]}[W_{3}W_{3}]
+\,\cC_{55}^{[35]}[W_{3}W_{5}]
+\,\cC_{55}^{[44]}[W_{4}W_{4}]
+\,\cC_{55}^{[33]''}\partial^{2}[W_{3}W_{3}]
\end{split}
\ee
Where all composite fields  can be calculated in the presence of stress-energy tensor $T(z)$ by thinking all multiple normal ordered products of
single primary  $W_{s}$ fields must be primary as in the eq.(2.2).The first two lines of ansatz (2.10) let us expression for composite primary operators
$[W_{3}W_{3}]$ , $[W_{3}W_{4}]$ and we find
\be
\begin{split}
[W_{3}W_{3}]\,
&=\,W_{3}W_{3}
 -\frac{22 \,\cC_{33}^{4}}{3 (c+24)} TW_{4}
 -\frac{(5 c+76) \,\cC_{33}^{4}}{36 (c+24)} \partial^{2}W_{4}\\
& -\frac{16 (191 c+22)}{3 (2 c-1) (5 c+22) (7 c+68)}TTT
 -\frac{225 c^2+1978 c+776}{2 (2 c-1) (5 c+22) (7 c+68)}\partial T\partial T\\
\end{split}
\ee
\be
\begin{split}
& -\frac{2 \left(67 c^2+178 c-752\right)}{(2 c-1) (5 c+22) (7 c+68)}\partial^{2}TT
 -\frac{(c-8) \left(5 c^2+60 c+4\right)}{6 (2 c-1) (5 c+22) (7 c+68)}\partial^{4}T\\
 \nonumber\\
\end{split}
\ee
and
\be
\begin{split}
[W_{3}W_{4}]\,
&=\,W_{3}W_{4}-\frac{94 \,\cC_{34}^{5}}{11 c+350} TR -\frac{4 (257 c+83)\, \cC_{34}^{3}}{(c+23) (5 c-4) (7 c+114)}TTW_{3} \\
&-\frac{\left(355 c^3-329 c^2-52214 c-12072\right)\, \cC_{34}^{3}}{18 (c+2) (c+23) (5 c-4) (7 c+114)}T\partial^{2}W_{3}
 -\frac{(5 c+22) \left(313 c^2+5783 c+2964\right)\, \cC_{34}^{3}}{36 (c+2) (c+23) (5 c-4) (7 c+114)}\partial T\partial T \\
&-\frac{\left(437 c^3+9089 c^2+22454 c-76152\right)\, \cC_{34}^{3}}{12 (c+2) (c+23) (5 c-4) (7 c+114)}\partial^{2}W_{3}
 -\frac{(c+19) \,\cC_{34}^{5}}{11 c+350}\partial^{2}W_{5} \\
&-\frac{\left(25 c^4-930 c^3-17157 c^2+115358 c+26904\right)\, \cC_{34}^{3}}{432 (c+2) (c+23) (5 c-4) (7 c+114)}\partial^{4}W_{3}\\
\end{split}
\ee
Where RHS of these equations contains all the descendants at given level. In spite of the fact that the coefficients look very complicated,
they are all established by the Virasoro algebra.Using function $\bf{OPEPPole}$ in the $\bf{OPEconf}$ packace one can find these results,
as well as all the fields appearing on the RHS of eq.(2.10). The first significant Jacobi identity is $(W_{3}W_{3}W_{4})$ which gives us
the first nontrivial relations for the structure constants appearing on the RHS of eq.(2.10).Then remainder Jacobi identities $(W_{3}W_{4}W_{4})$,
$(W_{4}W_{4}W_{4})$,$(W_{3}W_{4}W_{5})$,$(W_{3}W_{5}W_{5})$,$(W_{4}W_{5}W_{5})$ and $(W_{5}W_{5}W_{5})$ give us respectively
 all the structure constants appearing on the RHS of eq.(2.10).For completeness, the resulting relations are
\be
\begin{split}
&\cC_{33}^{4}\,=\,\sqrt{\frac{1024 (c+2) (c+23)}{3 (5 c+22) (7 c+68)}}\,,\,\,
 \cC_{34}^{3}\,=\,{3\over 4}\cC_{33}^{4}\,,\,\,\\
&\cC_{34}^{5}\,=\,\frac{960 (3 c+116)}{ (7 c+68) (7 c+114)\cC_{33}^{4} \cC_{45}^{[34]'} }\,=\,\sqrt{\frac{25 (3 c+116) (5 c+22)}{(7 c+68) (7 c+114)}}\,,\,\, \\
&\cC_{44}^{4}\,=\,\frac{96 \left(c^2+70 c-128\right)}{ (5 c+22) (7 c+68)\cC_{33}^{4}}\,=\,\sqrt{\frac{27 \left(c^2+70 c-128\right)^2}{(c+2) (c+23) (5 c+22) (7 c+68)}}\,,\,\,  \\
&\cC_{44}^{[33]}\,=\,\frac{9 (5 c+22)}{2 (c+2) (c+23)}\,,\,\,
 \cC_{35}^{4}\,=\, \frac{1}{48}  (5 c+22)\cC_{33}^{4} \cC_{45}^{[34]'} \,=\,\sqrt{\frac{16 (3 c+116) (5 c+22)}{(7 c+68) (7 c+114)}} \,,\,\,            \\
&\cC_{35}^{[33]}\,=\,\frac{2 (2 c-1) \cC_{45}^{[34]'}}{3 c+116}\,=\,\sqrt{\frac{432 (2 c-1)^2}{(c+2) (c+23) (3 c+116) (7 c+114)}} \,,\,\,
 \cC_{45}^{3}\,=\,\frac{3}{4}\cC_{35}^{4}\,,\,\,  \\
&\cC_{45}^{5}\,=\,-\frac{80 \left(11 c^3+204 c^2+9340 c+70272\right)}{ (5 c+22) (7 c+68) (7 c+114)\cC_{33}^{4}}\,=\,\sqrt{\frac{75 \left(11 c^3+204 c^2+9340 c+70272\right)^2}{4 (c+2) (c+23) (5 c+22) (7 c+68) (7 c+114)^2}}\,,\,\,\\
&\cC_{45}^{[34]}\,=\,\frac{(37 c+334) }{3 (3 c+116)}\cC_{45}^{[34]'}\,=\,\sqrt{\frac{12 (37 c+334)^2}{(c+2) (c+23) (3 c+116) (7 c+114)}}\,,\,\,
\cC_{55}^{4}\,=\,\frac{4}{5}\cC_{45}^{5}\,,\,\,\\
&\cC_{55}^{[33]}\,=\,\frac{3 \left(181 c^3+14880 c^2+248948 c+1507824\right)}{2 (c+2) (c+23) (3 c+116) (7 c+114)}\,,\,\,
 \cC_{55}^{[35]}\,=\,-\frac{10}{3}\cC_{45}^{[34]'}\,,\,\,\\
&\cC_{55}^{[44]}\,=\,\frac{64 (7 c+114)}{(3 c+116) (5 c+22)}\,,\,\,
 \left(\cC_{45}^{[34]'}\right)^{2}\,=\,\frac{108 (3 c+116)}{(c+2) (c+23) (7 c+114)}\,,\,\,\\
&\cC_{55}^{[33]''}\,=\,\frac{4 \left(11 c^2-306 c-13656\right)}{(2 + c) (116 + 3 c) (114 + 7 c)}.\\
\end{split}
\ee
These are also the same as in Ref. \cite{Hornfeck:1992tm}.

\section{The Quantum Miura transformation and the Casimir  ${\cW\cA}_{\it{5}}$ \,algebra}

The  $\cW_{\it{N}}$\, algebra is generated by a set of chiral currents $ \{ U_k (z)\} $ , of conformal dimension k$\,(k=1 , \cdots , N) $.\ Let us define a
differential  operator $ R_N (z) $ of degree ${N}$ \cite{Bouwknegt:1992wg}

\begin{equation}
R_{_N} (z)=-\sum_{k=0}^N U_{_k} (z) (\alpha_0 \partial)^{N-k}=: \prod_{j=1}^N \left(\alpha_{_0} \partial_{_z}- h_{_j}(z)  \right):,
\end{equation}

\noindent where  the double dots   denote the normal ordering of the free fields $\varphi(z)$.\
Here  $ \varphi(z) $ is an $ N-1 $ component Feigin Fuchs-type of free massless scalar fields .\ This transformation is called the quantum
Miura transformation and it determines completely the fields $\{U_{_k} (z)\}$ with
\begin{equation}
h_{_j}(z)=i  \mu_{_j}\partial \varphi(z)
\end{equation}
Here,\ ${\mu}_i $'s, $(i=1 , \cdots , N) $ are the weights of the fundamental representation of $ SU(N) $,~satisfying
$ \sum_{i=1}^N {\mu}_i=0 $ and ~${\mu}_i .{\mu}_j=\delta_{_{ij}}-{1\over N} $. The simple roots of
$ SU(N) $ are given by $ {\alpha}_i={\mu}_i-{\mu}_{i+1} $, $(i=1 , \cdots , N-1) $. \ The Weyl vector
of $ SU(N) $ is denoted as $ \rho={{1} \over {2}} \sum_{\alpha > 0} {\bf \alpha^{+}} $ where $\alpha^{+}$ are
the positive roots of $ SU(N) $.An OPE of $ h_{_j}(z)$ with itself is given by
\begin{equation}
h_{_i}(z_1)h_{_j}(z_2)\,\sim \,{{\delta_{_{ij}}-{1\over N}}\over {z_{12}}}
\end{equation}
The fields $ \{ U_k (z)\} $ can be obtained by expanding $R_{_N} (z)$.\ We present  a first five for $\cW_{\it{N}}$ algebra explicitely as in the following \cite{Ozer}
\be
\begin{split}
&U_{_0}(z)=-1\\
&U_{_1}(z)=\sum_{_i}\,h_{_i}(z)=0\\
&U_{_2}(z)=-\sum_{i<j}\,(h_{_i}\,h_{_j})(z)+\alpha_{_0}\,\sum_{_i}(i-1)\,\partial{h}_{_i}(z)\\
&U_{_3}(z)=\sum_{i<j<k}\,(h_{_i}\,h_{_j}\,h_{_k})(z)-\alpha_{_0}\,\sum_{_i<j}(i-1)\,\partial{\Big((h_{_i}\,h_{_j})(z)\Big)}\\
&-\alpha_{_0}\,\sum_{i<j}(j\!-\!i\!-\!1)\,(h_{_i}\,\partial{h}_{_j})(z)\!+\!{1\over 2} {\alpha_{_0}}^2 \sum_{i}(i\!-\!1)(i\!-\!2){\partial}^2{h}_{_i}(z)\\
\end{split}
\ee
\be
\begin{split}
&U_{_4}(z)=-\sum_{i<j<k<l}\,(h_{_i}\,h_{_j}\,h_{_k}\,h_{_l})(z)\!+\!\alpha_{_0}\sum_{_i<j<k}(i\!-\!1)({\partial}{h_{_i}}\,h_{_j}\,h_{_k})(z)\\
&+\alpha_{_0} \sum_{i<j<k}(j\!-\!2)\,(h_{_i}\,{\partial}{h}_{_j}h_{_k})(z)\!+\! \alpha_{_0} \sum_{i<j<k}(k\!-\!3)\,(h_{_i}\,h_{_j}{\partial}{h}_{_k})(z)\\
&+\alpha_{_0}^2\sum_{i<j}(i\!-\!1)(j\!-\!3)(\partial{h}_{_i}\,\partial{h}_{_j})(z)\!-\!{{\alpha_{_0}^2} \over 2}\sum_{i<j}(j\!-\!2)(j\!-\!3)(h_{_i}\,{\partial}^2{h}_{_j})(z)\\
&-{{\alpha_{_0}^2} \over 2} \sum_{i<j}(i\!-\!1)(i\!-\!2)({\partial}^2{h}_{_i}h_{_j})(z)+{{\alpha_{_0}^3} \over 3!}\sum_{i}(i\!-\!1)(i\!-\!2)(i\!-\!3){\partial}^3{h}_{_i}(z)\\
\end{split}
\ee
\be
\begin{split}
&U_{_5}(z)=\sum_{i<j<k<l<m}\,(h_{_i}\,h_{_j}\,h_{_k}\,h_{_l}h_m)(z)\\
&-\alpha_{_0}\sum_{_i<j<k<l}(i \,- \,1)(\partial{h_{_i}}\,h_{_j}\,h_{_k} h_{l})(z)-\alpha_{_0} \sum_{i<j<k<l}(j \,- \,2)\,(h_{_i}\,\partial{h_{_j}} h_{_k}h_{_l})(z)\\
&-\alpha_{_0} \sum_{i<j<k<l}(k \,- \,3)\,(h_{_i}\,h_{_j}\partial{h_{_k}} h_{_l})(z)-\alpha_{_0} \sum_{i<j<k<l}(l \,- \,4)\,(h_{_i}\,h_{_j} h_{_k}\partial{h_{_l}})(z)\\
&+{{\alpha_{_0}^2} \over 2}\sum_{i<j<k}(i \,- \,1)(i \,- \,2)({\partial}^2{h}_{_i}\,h_{_j}\,h_{_k})(z)\\
&+\,{{\alpha_{_0}^2} \over 2}\sum_{i<j<k}(j \,- \,2)(j \,- \,3)(h_{_i}\,{\partial}^2{h}_{_j} h_{_k})(z)+{{\alpha_{_0}^2} \over 2} \sum_{i<j<k}(k \,- \,3)(k \,- \,4)(h_{_i}h_{_j} {\partial}^2{h}_{_k})(z)\\
&+\alpha_{_0}^2\sum_{i<j<k}(i \,- \,1)(j \,- \,3)(\partial{h}_{_i}\,\partial{h}_{_j}h_{_k})(z) \,+ \, \alpha_{_0}^2\sum_{i<j<k}(i \,- \,1)(k \,- \,4)(\partial{h}_{_i}\,h_{_j} \partial{h}_{_k})(z)\\
&+\alpha_{_0}^2 \sum_{i<j<k}(j \,- \,2)(k \,- \,4)(h_{_i} \partial{h}_{_j}\partial{h}_{_k})(z)\\
&-{{\alpha_{_0}^3} \over 6}\sum_{i<j}(i \,- \,1)(i \,- \,2)(i \,- \,3)({\partial}^3{h}_{_i}h_{_j})(z)-{{\alpha_{_0}^3} \over 2}\sum_{i<j}(i \,- \,1)(i \,- \,2)(j \,- \,4)({\partial}^2{h}_{_i}h_{_j} h_{_k})(z)\\
&-{{\alpha_{_0}^3} \over 2}\sum_{i<j}(i \,- \,1)(j \,- \,3)(j \,- \,4)(\partial{h}_{_i}{\partial}^2{h}_{_j})(z)-{{\alpha_{_0}^3} \over 6}\sum_{i<j}(j \,- \,2)(j \,- \,3)(j \,- \,4)(h_{_i} {\partial}^3{h}_{_j})(z)\\
&+{{\alpha_{_0}}^4 \over 4!} \sum_{i}(i \,- \,1)(i \,- \,2)(i \,- \,3)(i \,- \,4){\partial}^4{h}_{_i}(z)\\
\end{split}
\ee
One can see that $ U_{_2} (z) \equiv T(z) $ has spin 2,~ which is called the stress-energy tensor,~ $ U_s (z) $ has spin s.
\ The standard OPE of $ T(z) $ with itself is
\be
\begin{split}
& T(z_1)T(z_2)\,\sim \,{{c\over 2}\over{z_{12}^{4}}}\,+\, {2\,T\over{z_{12}^{2}}}\, + \,{\partial T\over{z_{12}}}
\end{split}
\ee
where the central charge,~for $ SU(N) $,~is given by
\begin{equation}
c=(N-1)~(1-N(N+1) {\alpha_0}^2).
\end{equation}


\subsection{ Primary Field content of the Casimir ${\cW\cA}_{\it{5}}$ \,algebra in the Quantum Miura basis}

\noindent If one compare the fields $U_s (z)$ and the fields $W_{s}(z)$ as in the section-1, the fields $U_s (z)$ are not primary since \cite{Ozer}
\begin{equation}
T(z_1)U_{_k}(z_2)\,\sim \,{1 \over 2}\!\sum_{s=1}^k {(N\!-\!k\!+\!s)! \over (N\!-\!k)!}a_{_0}^{s\!-\!2}\Big(((s\!-\!1)(N\!-\!1)+2(k\!-\!1))a_{_0}^2\!-\!{{s\!-\!1}\over N}\Big){{U_{_{k\!-\!s}}} \over z_{12}^{s\!+\!2}}
+{{k U_{_k}} \over z_{12}^2}+{{\partial{U_{_k}}} \over z_{12}}
\end{equation}
Using above OPE,\ we want to construct the Casimir $\cW_{5}$\,algebra.\ Therefore we must obtain spin 3 , spin 4 and spin 5 primary fields. \ Here we first write
down the spin 3 primary field for $ SU(N) $ as \cite{Ozer}

\begin{equation}
\overline{U}_{_3}(z) =U_3 (z)-{(N-2) \over 2}~ \alpha_{_0}\partial{T}(z)
\end{equation}

the spin 4 primary field as \cite{Ozer}
\be
\begin{split}
\overline{U}_{_4} (z)= &U_{4} (z)-{(N-3) \over 2}\,a_{_0}\,~\partial U_{3} (z)\\
&+{(N-2)(N-3) \over {4N(22+5c)}}\,[-3+N(13+3N+2c)\,a_{_0}^2]\,~\partial^2 T(z)\\
&+{(N-2)(N-3) \over {2N(22+5c)}}\,[5-N(5N+7)\,a_{_0}^2]\,~(TT)(z)\\
\end{split}
\ee
and finally the spin 5 primary field as \cite{Ozer}
\be
\begin{split}
\overline{U}_{_5}(z)=& U_{_5}(z)-{(N-4) \over 2}\,a_{_0}\,\partial{U_{_4}}(z) \\
&+ {3 \over 4}\,{ (N-3)(N-4) \over {N(114+7c)}}\,[-2+N(20+C+2N)\,a_{_0}^2]\, \partial^{2} {U_{_3}}(z) \\
&+ {(N-2)(N-3)(N-4)\,a_{_0} \over {12N(114+7c)}}\,[9-N(33+C+9N)\,a_{_0}^2]\, \partial^{3} {U_{_2}}(z) \\
&+ {(N-3)(N-4) \over {N(114+7c)}}\,[7-N(13+7N)\,a_{_0}^2]\, (U_{_2}U_{_3})(z) \\
&+ { (N-2)(N-3)(N-4)\,a_{_0} \over {2N(114+7c)}}\,[-7+N(13+7N) \,a_{_0}^2]\, (U_{_2} \partial{U_{_2}})(z)\\
\end{split}
\ee


\subsection{ OPEs of higher spin primary fields}

To obtain  OPE of the two primary fields $\{\overline{U}_{_k} (z)\}$ and $\{\overline{U}_{_k \prime} (z)\}$ which
gives the central terms in the known form as
\be
\begin{split}
& W_{s}(z_1)W_{s}(z_2)\,\sim \,{{c\over s}\over{z_{12}^{2s}}}\,+\, \cdot\cdot\cdot
\end{split}
\ee
so, we must take care of the normalized forms of all the primary fields $\{\overline{U}_{_k} (z)\}$.\ Therefore the
normalized form of the Casimir $\cW_{\it{5}}$\,algebra generators are given by the following expressions:
\be
\begin{split}
\overline{U}_{_3} (z)=\sqrt{\theta_{3}} W_{3}(z)\, ,\, \overline{U}_{_4} (z)    =\sqrt{\theta_{4}} W_{4}(z)\, ,\,
\overline{U}_{_5} (z)    =\sqrt{\theta_{5}} W_{5}(z)
\end{split}
\ee

where $ \theta_{3} $ , $ \theta_{4}$ and $ \theta_{5}$ are the normalization factors for $ SU(5) $ and they can be written explicitly:
\begin{equation}
\theta_{_3}={{7 c+68} \over {80}}\, ,\, \theta_{4}={{(c+2) (c+23) (7 c+68)} \over {300 (5 c+22))}}\, ,\,\theta_{5}=\frac{(c+2) (c+23) (3 c+116) (7 c+68)}{24000 (7 c+114)}
\end{equation}

A straightforward calculation gives us the  non-trivial OPEs of the Casimir ${\cW\cA}_{\it{5}}$ \,algebra as in the Appendix, with structure constants which can be calculated explicitly:
\be
\begin{split}
&\cC_{33}^{4}\,=\, \frac{320}{(7 c+68) \theta _4}     \,=\,\sqrt{\frac{1024 (c+2) (c+23)}{3 (5 c+22) (7 c+68)}}\,,\,\,\\
&\cC_{34}^{3}\,=\,{3\over 4}\cC_{33}^{4}\,,\,\,
 \cC_{34}^{5}\,=\,\frac{5 \theta _3 \theta _4}{\theta _5}\,=\,\sqrt{\frac{25 (3 c+116) (5 c+22)}{(7 c+68) (7 c+114)}}\,,\,\, \\
&\cC_{44}^{4}\,=\,\frac{90 \left(c^2+70 c-128\right)}{(c+2) (c+23) (7 c+68) \theta _4}\,=\,\sqrt{\frac{27 \left(c^2+70 c-128\right)^2}{(c+2) (c+23) (5 c+22) (7 c+68)}} \,,\,\, \\
&\cC_{44}^{[33]}\,=\,\frac{9 (5 c+22)}{2 (c+2) (c+23)}\,,\,\,
 \cC_{35}^{4}\,=\, \frac{(3 c+116) (5 c+22) \theta _3 \theta _5}{20 (7 c+114) \theta _4} \,=\,\sqrt{\frac{16 (3 c+116) (5 c+22)}{(7 c+68) (7 c+114)}} \,,\,\,            \\
 \end{split}
\ee
\be
\begin{split}
&\cC_{35}^{[33]}\,=\,\frac{6 (2 c-1) \theta _5}{5 (7 c+114) \theta _3}\,=\,\sqrt{\frac{432 (2 c-1)^2}{(c+2) (c+23) (3 c+116) (7 c+114)}}\,,\,\,
 \cC_{45}^{3}\,=\,\frac{3}{4}\cC_{35}^{4} \,,\,\, \\
&\cC_{45}^{5}\,=\,-\frac{\left(11 c^3+204 c^2+9340 c+70272\right) \theta _4}{4 (5 c+22) (7 c+114)}\,=\,\sqrt{\frac{75 \left(11 c^3+204 c^2+9340 c+70272\right)^2}{4 (c+2) (c+23) (5 c+22) (7 c+68) (7 c+114)^2}}\,,\,\,\\
&\cC_{45}^{[34]}\,=\,\frac{(37 c+334) \theta _5}{5 (7 c+114) \theta _3}\,=\,\sqrt{\frac{12 (37 c+334)^2}{(c+2) (c+23) (3 c+116) (7 c+114)}}\,,\,\,
\cC_{55}^{4}\,=\,\frac{4}{5}\cC_{45}^{5}\,,\,\,\\
&\cC_{55}^{[33]}\,=\, \frac{120 \left(181 c^3+14880 c^2+248948 c+1507824\right)}{(c+2) (c+23) (3 c+116) (7 c+68) (7 c+114) \theta _3^2} \,=\,\frac{3 \left(181 c^3+14880 c^2+248948 c+1507824\right)}{2 (c+2) (c+23) (3 c+116) (7 c+114)}\,,\,\,\\
&\cC_{55}^{[35]}\,=\, -\frac{2 (3 c+116) \theta _5}{(7 c+114) \theta _3}  \,=\,-\frac{10}{3}\cC_{45}^{[34]'}\,,\,\,\left(\cC_{45}^{[34]'}\right)^{2}\,=\,\frac{108 (3 c+116)}{(c+2) (c+23) (7 c+114)}\,,\,\,\\
&\cC_{55}^{[44]}\,=\, \frac{19200 (7 c+114)}{(c+2) (c+23) (3 c+116) (7 c+68) \theta _4^2} \,=\,\frac{64 (7 c+114)}{(3 c+116) (5 c+22)}.
\nonumber\\
\end{split}
\ee
These structure constants  overlap with the structure constants(2.13) in the Casimir ${\cW\cA}_{\it{N}}$\,algebras  in the section 2.
\newpage
\appendix
\section{Appendix}
\subsection{Explicit OPEs of the  Casimir ${\cW\cA}_{\it{5}}$\,algebras }
Here we give all the non-trivial OPEs of the Casimir ${\cW\cA}_{\it{5}}$ \,algebra,as follows:
\be
\begin{split}& W_{3}(z_1)W_{3}(z_2)\,\sim \,
{{c\over 3}\over{z_{12}^{6}}}
 \,+\, {2\,T\over{z_{12}^{4}}}
 \,+\,{\partial T\over{z_{12}^{3}}}\\
& + {{1}\over{z_{12}^{2}}}\Big(
  \cC_{33}^{4} W_{4}
  + \frac {32} {5 c + 22} TT
  + \frac {3 (c - 2)} {2 (5 c + 22)}\partial^2 T
\Big)
 + {{1}\over{z_{12}^{ }}}\Big(
   \frac{\cC_{33}^{4}}{2}\partial W_{4}
 +\frac{32}{5 c+22} \partial T T
 +\frac{c-2}{3 (5 c+22)}\partial^3 T
\Big)\\
\end{split}
\ee
\be
\begin{split}
& W_{3}(z_1)W_{4}(z_2)\,\sim \,
  + {{1}\over{z_{12}^{4}}}
  \frac{3 \cC_{33}^{4}}{4} W_{3}
 + {{1}\over{z_{12}^{3}}}
\frac{\cC_{33}^{4}}{4} \partial W_{3}
+ {{1}\over{z_{12}^{2}}}\Big(
   \cC_{34}^{5}W_{5}
 +\frac{39 \cC_{33}^{4}}{7 c+114} T W_{3}
 +\frac{3 (c-6) \cC_{33}^{4}}{8 (7 c+114)}\partial^2 W_{3}
\Big)\\
& + {{1}\over{z_{12}^{ }}}\Big(
   \frac{2}{5}\cC_{34}^{5}\partial W_{5}
  + \frac{3 (9 c-2) \cC_{33}^{4}}{2 (c+2) (7 c+114)} T \partial W_{3}
  + \frac{15 (5 c+22) \cC_{33}^{4}}{4 (c+2) (7 c+114)}\partial T W_{3}
 +\frac{\left(c^2-31 c-6\right) \cC_{33}^{4}}{16 (c+2) (7 c+114)}\partial^3 W_{3}
\Big)\\
\end{split}
\ee
\be
\begin{split}& W_{4}(z_1)W_{4}(z_2)\,\sim \,
{{c\over 4}\over{z_{12}^{8}}}
 \,+\, {2\,T\over{z_{12}^{6}}}
 \,+\,{\partial T\over{z_{12}^{5}}}\\
& + {{1}\over{z_{12}^{4}}}\Big(
   \cC_{44}^{4} W_{4}
  +\frac{42}{5 c+22} TT
  +\frac{3 (c-4)}{2 (5 c+22)}\partial^2  T
\Big)
 + {{1}\over{z_{12}^{3}}}\Big(
  \frac{\cC_{44}^{4}}{2}\partial   W_{4}
 +\frac{42}{5 c+22} \partial   T   T
 +\frac{c-4}{3 (5 c+22)} \partial^3  T
\Big)\\
& + {{1}\over{z_{12}^{2}}}\Big(
  \cC_{44}^{[33]} W_{3}W_{3}
 +\frac{128 (7 c-118)}{(5 c+22) (7 c+68) \cC_{33}^{4}} TW_{4}
 +\frac{72 (3 c+38) (4 c-1)}{(c+2) (c+23) (5 c+22) (7 c+68)} TTT\\
 &+\frac{295 c^3+5052 c^2+16164 c-34768}{4 (c+2) (c+23) (5 c+22) (7 c+68)} \partial   T    \partial   T
 +\frac{88 c^3+795 c^2-5994 c-32608}{(c+2) (c+23) (5 c+22) (7 c+68)} \partial^2  T   T  \\
 &+\frac{8 \left(5 c^2-106 c-1456\right)}{3 (5 c+22) (7 c+68) \cC_{33}^{4}} \partial^2  W_{4}
 +\frac{5 c^4+20 c^3-1344 c^2-10928 c+4336}{12 (c+2) (c+23) (5 c+22) (7 c+68)} \partial^4  T
\Big)\\
& + {{1}\over{z_{12}^{ }}}\Big(
   \cC_{44}^{[33]} \partial   W_{3}   W_{3}
  +\frac{64 (7 c-118)}{(5 c+22) (7 c+68) \cC_{33}^{4}}   T   \partial   W_{4}
  +\frac{64 (7 c-118)}{(5 c+22) (7 c+68) \cC_{33}^{4}}   \partial   T   W_{4}  \\
 & +\frac{108 (3 c+38) (4 c-1)}{(c+2) (c+23) (5 c+22) (7 c+68)} \partial   T     T  T
 +\frac{3 \left(59 c^3+492 c^2-3204 c-5312\right)}{4 (c+2) (c+23) (5 c+22) (7 c+68)}\partial^2  T    \partial   T \\
 &+\frac{3 (c-8) \left(13 c^2+226 c+832\right)}{2 (c+2) (c+23) (5 c+22) (7 c+68)}  \partial^3  T   T
 +\frac{8 \left(c^2-38 c-8\right)}{3 (5 c+22) (7 c+68) \cC_{33}^{4}} \partial^3  W_{4}\\
 &+\frac{5 c^4+20 c^3-1344 c^2-10928 c+4336}{80 (c+2) (c+23) (5 c+22) (7 c+68)} \partial^5  T
\Big) \\
\end{split}
\ee
\be
\begin{split}
& W_{3}(z_1)W_{5}(z_2)\,\sim \,
  {{1}\over{z_{12}^{4}}}
 \cC_{35}^{4}W_{4}
 + {{1}\over{z_{12}^{3}}}
\frac{\cC_{35}^{4}}{4} \partial W_{4}\\
& + {{1}\over{z_{12}^{2}}}\Big(
  \frac{5 (3 c+22) \cC_{33}^{4} \cC_{45}^{[34]'}}{6 (3 c+116)} TW_{4}
 -\frac{32 (191 c+22) \cC_{45}^{[34]'}}{3 (3 c+116) (5 c+22) (7 c+68)} TTT
 +\frac{2 (2 c-1) \cC_{45}^{[34]'}}{3 c+116}W_{3}W_{3}\\
 &-\frac{\left(225 c^2+1978 c+776\right) \cC_{45}^{[34]'}}{(3 c+116) (5 c+22) (7 c+68)} \partial T \partial T
 -\frac{4 \left(67 c^2+178 c-752\right) \cC_{45}^{[34]'}}{(3 c+116) (5 c+22) (7 c+68)}\partial^2T T\\
 &+\frac{\left(5 c^2-158 c-216\right) \cC_{33}^{4} \cC_{45}^{[34]'}}{384 (3 c+116)} \partial^2W_{4}
 -\frac{(c-8) \left(5 c^2+60 c+4\right) \cC_{45}^{[34]'}}{3 (3 c+116) (5 c+22) (7 c+68)}\partial^4T
\Big)\nonumber\\
\end{split}
\ee
\be
\begin{split}
& + {{1}\over{z_{12}^{ }}}\Big(
 \frac{(2 c-1) \cC_{33}^{4} \cC_{45}^{[34]'}}{3 (3 c+116)} T \partial W_{4}
 +\frac{(7 c+114) \cC_{33}^{4} \cC_{45}^{[34]'}}{6 (3 c+116)} \partial T W_{4}
 -\frac{32 (191 c+22) \cC_{45}^{[34]'}}{3 (3 c+116) (5 c+22) (7 c+68)}\partial T T T \\
 &+\frac{4 (2 c-1) \cC_{45}^{[34]'}}{3 (3 c+116)}\partial W_{3} W_{3}
 -\frac{30 \left(3 c^2+6 c+8\right) \cC_{45}^{[34]'}}{(3 c+116) (5 c+22) (7 c+68)}\partial^2T \partial T
 -\frac{4 \left(89 c^2+266 c-1624\right) \cC_{45}^{[34]'}}{9 (3 c+116) (5 c+22) (7 c+68)}\partial^3T T \\
 &+\frac{\left(c^2-70 c-24\right) \cC_{33}^{4} \cC_{45}^{[34]'}}{576 (3 c+116)} \partial^3W_{4}
 -\frac{(c-8) \left(5 c^2+60 c+4\right) \cC_{45}^{[34]'}}{30 (3 c+116) (5 c+22) (7 c+68)}\partial^5T
\Big)\\
\end{split}
\ee
\be
\begin{split}
& W_{4}(z_1)W_{5}(z_2)\,\sim \,
   {{1}\over{z_{12}^{6}}}
\cC_{45}^{3}W_{3}
 + {{1}\over{z_{12}^{5}}}
\frac {\cC_{45}^{3}} {3} \partial W_{3}\\
& + {{1}\over{z_{12}^{4}}}\Big(
 \cC_{45}^{5} W_{5}
 +\frac {33 (5 c+22) \cC_{33}^{4} \cC_{45}^{[34]'}}{32 (7 c+114)}TW_{3}
 +\frac {(c-12) (5 c+22) \cC_{33}^{4} \cC_{45}^{[34]'}}{128 (7 c+114)}\partial^  2W_{3}
\Big)\\
& + {{1}\over{z_{12}^{3}}}\Big(
  \frac {2}{5}  \cC_{45}^{5}\partial  W_{5}
  +\frac {(5 c+22) (17 c-6) \cC_{33}^{4} \cC_{45}^{[34]'}}{48 (c+2) (7 c+114)} T  \partial  W_{3}\\
 & +\frac {(2 c+9) (5 c+22) \cC_{33}^{4} \cC_{45}^{[34]'}}{4 (c+2) (7 c+114)} \partial  T  W_{3}
 +\frac {(5 c+22) \left(c^2-44 c-12\right) \cC_{33}^{4} \cC_{45}^{[34]'}}{768 (c+2) (7 c+114)}\partial^  3W_{3}
\Big)\\
 &+ {{1}\over{z_{12}^{2}}}\Big(
   \cC_{45}^{[34]}  W_{3}  W_{4}
 -\frac {320 \left(29 c^2+1196 c+7796\right)}{(5 c+22) (7 c+68) (7 c+114) \cC_{33}^{4}}T  W_{5}
 +\frac {\left(1224 c^2+23921 c-28834\right) \cC_{33}^{4} \cC_{45}^{[34]'}}{16 (c+23) (3 c+116) (7 c+114)} T  T  W_{3}\\
 &+\frac {\left(445 c^4+2354 c^3-255316 c^2-2192552 c+1687872\right) \cC_{33}^{4} \cC_{45}^{[34]'}}{384 (c+2) (c+23) (3 c+116) (7 c+114)}T  \partial^ 2W_{3} \\
 &+\frac {(5 c+22) \left(397 c^3+12392 c^2+66564 c-333264\right) \cC_{33}^{4} \cC_{45}^{[34]'}}{768 (c+2) (c+23) (3 c+116) (7 c+114)}\partial  T  \partial  W_{3} \\
 &+\frac {\left(563 c^4+18130 c^3+73748 c^2-1597032 c-8393024\right) \cC_{33}^{4} \cC_{45}^{[34]'}}{256 (c+2) (c+23) (3 c+116) (7 c+114)} \partial^  2T  W_{3}
 -\frac {80 \left(c^3+70 c^2+1100 c+4808\right)}{(5 c+22) (7 c+68) (7 c+114) \cC_{33}^{4}}\partial^  2W_{5} \\
 &+\frac {\left(25 c^5-1325 c^4-49678 c^3+92572 c^2+6272584 c-3320544\right) \cC_{33}^{4} \cC_{45}^{[34]'}}{9216 (c+2) (c+23) (3 c+116) (7 c+114)}\partial^  4W_{3}
\Big)\nonumber\\
\end{split}
\ee
\be
\begin{split}
& + {{1}\over{z_{12}^{}}}\Big(
 \frac {2 (2 c-1) \cC_{45}^{[34]'}}{3 c+116} W_{3} \partial  W_{4}
  +\frac {(7 c+114) \cC_{45}^{[34]'}}{3 c+116}\partial  W_{3} W_{4}
 -\frac {64 \left(59 c^2+2488 c+12900\right)}{(5 c+22) (7 c+68) (7 c+114) \cC_{33}^{4}} T \partial  W_{5} \\
 &-\frac {640 (c+23)}{(5 c+22) (7 c+68) \cC_{33}^{4}}\partial  T W_{5}
 +\frac {3 \left(144 c^3+1871 c^2-24304 c-2140\right) \cC_{33}^{4} \cC_{45}^{[34]'}}{16 (c+2) (c+23) (3 c+116) (7 c+114)}T T \partial  W_{3} \\
 &+\frac {\left(25 c^4-1648 c^3-46548 c^2+32144 c+164896\right) \cC_{33}^{4} \cC_{45}^{[34]'}}{128 (c+2) (c+23) (3 c+116) (7 c+114)} T \partial^ 3W_{3}\\
 &+\frac {3 (2 c-1) \left(99 c^2+2644 c+12812\right) \cC_{33}^{4} \cC_{45}^{[34]'}}{8 (c+2) (c+23) (3 c+116) (7 c+114)}\partial  T T W_{3} \\
 &+\frac {\left(145 c^4-5518 c^3-216412 c^2-543272 c-180960\right) \cC_{33}^{4} \cC_{45}^{[34]'}}{256 (c+2) (c+23) (3 c+116) (7 c+114)}\partial  T \partial^ 2W_{3} \\
 &+\frac {\left(97 c^4-1212 c^3-82272 c^2-158800 c-172560\right) \cC_{33}^{4} \cC_{45}^{[34]'}}{128 (c+2) (c+23) (3 c+116) (7 c+114)} \partial^2T \partial  W_{3}\\
 \nonumber\\
\end{split}
\ee
\be
\begin{split}
  &+\frac {\left(123 c^4+1518 c^3-64036 c^2-893144 c-2412768\right) \cC_{33}^{4} \cC_{45}^{[34]'}}{256 (c+2) (c+23) (3 c+116) (7 c+114)}\partial^ 3T W_{3}\\
 &-\frac {8 \left(c^2+122 c-120\right)}{3 (7 c+68) (7 c+114) \cC_{33}^{4}} \partial^ 3W_{5}
 +\frac {\left(5 c^5-829 c^4+5194 c^3+628220 c^2+938024 c-1877856\right) \cC_{33}^{4} \cC_{45}^{[34]'}}{15360 (c+2) (c+23) (3 c+116) (7 c+114)}\partial^ 5W_{3}
\Big)\\
\end{split}
\ee

\be
\begin{split}
& W_{5}(z_1)W_{5}(z_2)\,\sim \,
{{c\over 5}\over{z_{12}^{10}}}
 \,+\, {2\,T\over{z_{12}^{8}}}
 \,+\,{\partial T\over{z_{12}^{7}}}\\
& + {{1}\over{z_{12}^{6}}}\Big(
  \cC_{55}^{4} W_{4}
 +\frac{52}{5 c+22} TT
 +\frac{3 (c-6)}{2 (5 c+22)}\partial^2 T
\Big)
 + {{1}\over{z_{12}^{5}}}\Big(
\frac{1}{2}\cC_{55}^{4} \partial W_{4}
+\frac{52}{5 c+22} \partial TT
+\frac{c-6}{3 (5 c+22)}\partial^3 T
\Big)\\
& + {{1}\over{z_{12}^{4}}}\Big(
 \cC_{44}^{[33]} W_{3}W_{3}
-\frac  {3 \left(187 c^3+11520 c^2+452876 c+3767568\right) \cC_{33}^{4}}{8 (c+2) (c+23) (3 c+116) (7 c+114)} TW_{4}\\
& +\frac  {24 \left(1148 c^4+86853 c^3+1942364 c^2+14490156 c-3744688\right)}{(c+2) (c+23) (3 c+116) (5 c+22) (7 c+68) (7 c+114)} TTT\\
& +\frac  {7665 c^5+622906 c^4+15925036 c^3+164020888 c^2+433533312 c-1132361856}{4 (c+2) (c+23) (3 c+116) (5 c+22) (7 c+68) (7 c+114)}\partial T\partial T \\
& +\frac  {2289 c^5+162920 c^4+2945488 c^3+5058544 c^2-270226960 c-1077929664}{(c+2) (c+23) (3 c+116) (5 c+22) (7 c+68) (7 c+114)}\partial^2TT \\
& -\frac  {\left(55 c^4+4812 c^3+238460 c^2+3664896 c+18619968\right) \cC_{33}^{4}}{64 (c+2) (c+23) (3 c+116) (7 c+114)}\partial^2W_{4} \\
& +\frac  {105 c^6+6880 c^5+61888 c^4-2504072 c^3-57614256 c^2-331490784 c+161834112}{12 (c+2) (c+23) (3 c+116) (5 c+22) (7 c+68) (7 c+114)}\partial^4T
\Big)\\
& + {{1}\over{z_{12}^{3}}}\Big(
   \cC_{44}^{[33]}\partial   W_{3}W_{3}
   -\frac  {3 \left(187 c^3+11520 c^2+452876 c+3767568\right) \cC_{33}^{4}}{16 (c+2) (c+23) (3 c+116) (7 c+114)} T\partial   W_{4}\\
 &-\frac  {3 \left(187 c^3+11520 c^2+452876 c+3767568\right) \cC_{33}^{4}}{16 (c+2) (c+23) (3 c+116) (7 c+114)} \partial   TW_{4}\\
 & +\frac  {36 \left(1148 c^4+86853 c^3+1942364 c^2+14490156 c-3744688\right)}{(c+2) (c+23) (3 c+116) (5 c+22) (7 c+68) (7 c+114)} \partial   TTT\\
  & +\frac  {3 \left(1533 c^5+108050 c^4+1934324 c^3+4834136 c^2-121951584 c-172548864\right)}{4 (c+2) (c+23) (3 c+116) (5 c+22) (7 c+68) (7 c+114)}\partial^2T\partial   T \\
 & +\frac  {3045 c^5+217790 c^4+3956652 c^3+5282952 c^2-418502336 c-1983310464}{6 (c+2) (c+23) (3 c+116) (5 c+22) (7 c+68) (7 c+114)}\partial^3TT \\
 & -\frac  {\left(11 c^4+738 c^3+33868 c^2+189528 c-797088\right) \cC_{33}^{4}}{64 (c+2) (c+23) (3 c+116) (7 c+114)} \partial^3W_{4}\\
 & +\frac  {105 c^6+6880 c^5+61888 c^4-2504072 c^3-57614256 c^2-331490784 c+161834112}{80 (c+2) (c+23) (3 c+116) (5 c+22) (7 c+68) (7 c+114)}\partial^5T
 \Big)\\
& + {{1}\over{z_{12}^{2}}}\Big(
 \cC_{55}^{[44]}  W_{4}W_{4}
+\cC_{55}^{[35]}W_{3}W_{5}
-\frac{16384 \left(43 c^3+2393 c^2+23131 c-5266\right)}{(3 c+116) (5 c+22)^2 (7 c+68) (7 c+114) \cC_{33}^{4}}TTW_{4}  \\
&+\frac{768 \left(504 c^4+17652 c^3+171793 c^2-84704 c+10972\right)}{(c+2) (c+23) (3 c+116) (5 c+22)^2 (7 c+68) (7 c+114)} TTTT
 +\frac{48 (2 c-1) (31 c+572)}{(c+2) (c+23) (3 c+116) (7 c+114)} TW_{3}W_{3} \\
&-\frac{32 \left(1555 c^4+163090 c^3+4385068 c^2+32825112 c-26037408\right)}{3 (3 c+116) (5 c+22)^2 (7 c+68) (7 c+114) \cC_{33}^{4}} T\partial^2W_{4} \\
&+\frac{4 \left(25900 c^5+1335057 c^4+22631046 c^3+117504164 c^2-223997112 c+48876576\right)}{(c+2) (c+23) (3 c+116) (5 c+22)^2 (7 c+68) (7 c+114)} \partial T\partial TT  \\
\nonumber\\
\end{split}
\ee
\be
\begin{split}
&-\frac{128 \left(235 c^4+16593 c^3+444596 c^2+2945348 c-2562704\right)}{(3 c+116) (5 c+22)^2 (7 c+68) (7 c+114) \cC_{33}^{4}}  \partial T\partial W_{4} \\
&+\frac{245 c^3+8536 c^2+31524 c-259536}{4 (c+2) (c+23) (3 c+116) (7 c+114)}  \partial W_{3}\partial W_{3}
+\frac{3 \left(111 c^3+2704 c^2-44676 c-924080\right)}{4 (c+2) (c+23) (3 c+116) (7 c+114)} \partial^2W_{3}W_{3}   \\
&-\frac{32 \left(557 c^4+23830 c^3+223380 c^2-10662136 c-105450464\right)}{(3 c+116) (5 c+22)^2 (7 c+68) (7 c+114) \cC_{33}^{4}}  \partial^2TW_{4} \\
&+\frac{2 \left(30884 c^5+1049739 c^4+4568478 c^3-161344916 c^2-1352580792 c+345598368\right)}{(c+2) (c+23) (3 c+116) (5 c+22)^2 (7 c+68) (7 c+114)} \partial^2TTT  \\
&+\frac{3 \left(2303 c^6+79492 c^5-477096 c^4-30376928 c^3-111113680 c^2+1333008960 c+277883904\right)}{8 (c+2) (c+23) (3 c+116) (5 c+22)^2 (7 c+68) (7 c+114)}\partial^2T\partial^2T \\
&+\frac{30625 c^6+1406272 c^5+4388712 c^4-432405472 c^3-4946700400 c^2-8004630912 c+12394720512}{24 (c+2) (c+23) (3 c+116) (5 c+22)^2 (7 c+68) (7 c+114)} \partial^3T\partial T  \\
&+\frac{5425 c^6+160666 c^5-3951116 c^4-170672872 c^3-1654958176 c^2-2594536192 c+5584883712}{12 (c+2) (c+23) (3 c+116) (5 c+22)^2 (7 c+68) (7 c+114)} \partial^4TT  \\
&-\frac{4 \left(35 c^5+3270 c^4+87668 c^3-775704 c^2-23503552 c-2991744\right)}{3 (3 c+116) (5 c+22)^2 (7 c+68) (7 c+114) \cC_{33}^{4}}  \partial^4W_{4}\\
&+\frac{1225 c^7+30940 c^6-2227980 c^5-77243512 c^4-744781216 c^3-1705759584 c^2+7306651392 c-2606286336}{1440 (c+2) (c+23) (3 c+116) (5 c+22)^2 (7 c+68) (7 c+114)} \partial^6T
 \Big)\\
& + {{1}\over{z_{12}}}\Big(
  \cC_{55}^{[44]}   \partial W_{4}  W_{4}
 + \cC_{55}^{[35]}  \partial (W_{3} W_{5})
 -\frac{8192 \left(43 c^3+2393 c^2+23131 c-5266\right)}{(3 c+116) (5 c+22)^2 (7 c+68) (7 c+114) \cC_{33}^{4}}T  T  \partial W_{4}  \\
&+\frac{48 (2 c-1) (31 c+572)}{(c+2) (c+23) (3 c+116) (7 c+114)} T  \partial W_{3}  W_{3}
 -\frac{64 \left(155 c^4+9936 c^3+158844 c^2+1871664 c-2291840\right)}{3 (3 c+116) (5 c+22)^2 (7 c+68) (7 c+114) \cC_{33}^{4}}  T  \partial^3W_{4}\\
& -\frac{16384 \left(43 c^3+2393 c^2+23131 c-5266\right)}{(3 c+116) (5 c+22)^2 (7 c+68) (7 c+114) \cC_{33}^{4}}\partial T  T  W_{4}
 + \frac{24 (2 c-1) (31 c+572)}{(c+2) (c+23) (3 c+116) (7 c+114)}  \partial T  W_{3}  W_{3} \\
&+ \frac{1536 \left(504 c^4+17652 c^3+171793 c^2-84704 c+10972\right)}{(c+2) (c+23) (3 c+116) (5 c+22)^2 (7 c+68) (7 c+114)} \partial T  T  T  T  \\
&+\frac{4 \left(868 c^4+25699 c^3+186946 c^2-1053888 c+262944\right)}{(c+2) (c+23) (3 c+116) (5 c+22) (7 c+68) (7 c+114)} \partial T  \partial T  \partial T  \\
& -\frac{32 \left(157 c^3+10430 c^2+67296 c-250272\right)}{3 (3 c+116) (5 c+22) (7 c+68) (7 c+114) \cC_{33}^{4}}  \partial T  \partial^2W_{4} \\
&+ \frac{8 \left(7756 c^5+254217 c^4+1074504 c^3-41564620 c^2-248593104 c+41783424\right)}{(c+2) (c+23) (3 c+116) (5 c+22)^2 (7 c+68) (7 c+114)} \partial^2T  \partial T  T \\
& -\frac{32 \left(281 c^4+3268 c^3-430508 c^2-10725536 c-50220992\right)}{(3 c+116) (5 c+22)^2 (7 c+68) (7 c+114) \cC_{33}^{4}} \partial^2T  \partial W_{4}  \\
&+ \frac{35 c^3+1096 c^2+12684 c-41136}{(c+2) (c+23) (3 c+116) (7 c+114)}  \partial^2W_{3}  \partial W_{3}
 +\frac{(c-24) \left(19 c^2+786 c+9296\right)}{(c+2) (c+23) (3 c+116) (7 c+114)}  \partial^3W_{3}  W_{3} \\
& -\frac{256 \left(46 c^4-73 c^3-98752 c^2-3039700 c-18902512\right)}{3 (3 c+116) (5 c+22)^2 (7 c+68) (7 c+114) \cC_{33}^{4}}  \partial^3T  W_{4} \\
&+ \frac{8 \left(1708 c^5+60145 c^4+77830 c^3-16938364 c^2-128486776 c+25371680\right)}{(c+2) (c+23) (3 c+116) (5 c+22)^2 (7 c+68) (7 c+114)} \partial^3T  T  T  \\
&+ \frac{2303 c^6+79492 c^5-882096 c^4-48628928 c^3-215981680 c^2+1148688960 c+94139904}{6 (c+2) (c+23) (3 c+116) (5 c+22)^2 (7 c+68) (7 c+114)} \partial^3T  \partial^2T  \\
 \nonumber\\
\end{split}
\ee
\be
\begin{split}
&+ \frac{1365 c^6+37234 c^5-1033716 c^4-44429640 c^3-386767936 c^2-483555456 c+910903296}{6 (c+2) (c+23) (3 c+116) (5 c+22)^2 (7 c+68) (7 c+114)} \partial^4T  \partial T  \\
&+ \frac{1015 c^6+30858 c^5-929600 c^4-40783408 c^3-373238160 c^2-472204384 c+1790162304}{15 (c+2) (c+23) (3 c+116) (5 c+22)^2 (7 c+68) (7 c+114)}  \partial^5T  T \\
&-\frac{8 \left(35 c^5+1590 c^4+48252 c^3+422632 c^2-25026336 c+13821696\right)}{45 (3 c+116) (5 c+22)^2 (7 c+68) (7 c+114) \cC_{33}^{4}} \partial^5W_{4} \\
&+ \frac{245 c^7+6188 c^6-592564 c^5-21763712 c^4-204399584 c^3-340348544 c^2+2548955904 c-556655616}{2520 (c+2) (c+23) (3 c+116) (5 c+22)^2 (7 c+68) (7 c+114)} \partial^7T
 \Big)\\
\end{split}
\ee


\acknowledgments
I would like to thank to Tom\'{a}\v{s} Proch\'{a}zka for useful discussions about the use of the Mathematica package $\bf{OPEconf}$
an also to the owner of the package Chris Thielemans since  this calculations can not be performed  easily without this package.


\end{document}